\documentclass[aps,prx,twocolumn,showpacs,longbibliography]{revtex4-1}
\usepackage[dvips]{graphicx}          
\usepackage{amsfonts,amssymb,wasysym} 
\begin{document}
\title{Ten reasons why a thermalized system cannot be described by a many-particle wave function}
\author{Barbara Drossel} 
\affiliation{Institut f\"ur Festk\"orperphysik,  TU Darmstadt,
Hochschulstra\ss e 6, 64289 Darmstadt, Germany; email: drossel@fkp.tu-darmstadt.de }
\date{\today}

\begin{abstract}
It is widely believed that the underlying reality behind statistical mechanics is a deterministic and unitary time evolution of a many-particle wave function, even though this is in conflict with the irreversible, stochastic nature of statistical mechanics. The usual attempts to resolve this conflict for instance by appealing to decoherence or eigenstate thermalization are riddled with problems. This paper considers theoretical physics of thermalized systems as it is done in practise and shows that all approaches to thermalized systems presuppose in some form limits to linear superposition and deterministic time evolution. These considerations include, among others,  the classical limit,  extensivity, the concepts of entropy and equilibrium, and symmetry breaking in phase transitions and quantum measurement. As a conclusion, the paper argues that the irreversibility and stochasticity of statistical mechanics should be taken as a true property of nature.  It follows that a gas of a macroscopic number $N$ of atoms in thermal equilibrium is best represented by a collection of $N$ wave packets of a size of the order of the thermal de Broglie wave length, which behave quantum mechanically below this scale but classically sufficiently far beyond this scale. In particular, these wave packets must localize again after scattering events, which requires stochasticity and indicates a connection to the measurement process.  
\end{abstract}
\maketitle

\section{Introduction}

Connecting quantum statistical mechanics (QSM) to many-particle nonrelativistic quantum mechanics (QM) poses severe conceptual problems since quantum mechanics is a deterministic theory for pure quantum states, while statistical mechanics (SM) is based on the concept of probabilities and uses mixed states. Despite of these essential differences between the two types of theories, most publications in the field of the foundations of quantum statistical mechanics \cite{Emch2007-EMCQSP,lindenpopescuetal2010, ShortFarrelly2012, reimann2013quantum,eisert2015quantum,gogolin2016equilibration} aim at "deriving" the properties of thermalized many-particle systems by starting from the $N$-particle Schr\"odinger equation 
\begin{equation}
i\hbar \frac {\partial \psi(\vec x_1,\dots,\vec x_N) }{\partial t} = H\psi(\vec x_1,\dots,\vec x_N)
\end{equation}
with 
\begin{equation}
\hat H= \sum_{\alpha=1}^N \frac{\hat p_\alpha^2}{2m_\alpha} + \frac 1 2\sum_{\alpha\neq\beta}V(\vec r_\alpha - \vec r_\beta) \, .\label{fullhamiltonian}
\end{equation}
 A promising concept to achieve this goal for isolated systems is the eigenstate thermalization hypothesis \cite{deutsch1991quantum,srednicki1994chaos}, which is based on the idea that the finite-energy eigenfunctions of interacting many-particle systems are "random" in a suitable sense, such that the expectation values of thermodynamic observables calculated with these eigenfunctions are identical to the expectation values obtained from statistical mechanics. A more comprehensive research agenda treats thermalized systems as open systems that are embedded in a larger environment. The combined system consisting of the considered system and the environment is taken to be isolated and is modeled by a many-particle Schr\"odinger equation.   Due to the interaction with the environment, which imposes the temperature on the system, the system becomes entangled with the environment. When the trace over the environmental degrees is taken, one obtains the reduced density matrix of the system, and this density matrix is that of a mixed state. This mixed state becomes that of quantum statistical mechanics when the non-diagonal elements vanish in the basis of the energy eigenfunctions. This is achieved by making plausible assumptions about the uncorrelatedness of the degrees of freedom of the environment.  Similar arguments can be applied to small subvolumes of a large isolated system: due to the interaction with the rest of the system, this subvolume becomes entangled with it, and its reduced density matrix is that of a mixed state. The combined system remains in a pure state and retains all information about the initial state \cite{popescu2006entanglement}. 

These promising results, combined with the impressive empirical successes at creating many-particle entanglements and quantum superpositions of mesoscopic objects, lead to the widespread belief that at the microscopic level nature shares the fundamental features of quantum mechanics, including a deterministic, reversible time evolution, linear superposition, and a huge amount of entanglement of all particles that have interacted with each other. 

However, this approach brings severe problems with it. The first problem is that of interpreting the mixed state of a thermalized system. While the Schr\"odinger equation for the combined system is usually interpreted as describing one system (and not an ensemble), the mixed state is taken to represent an ensemble of systems. This problem is analogous to the problem of interpreting the measurement process, where a unitary time evolution according to the Schr\"odinger equation leads to a mixed state of the observed particle, representing a superposition of all measurement outcomes and not to just one of them, as observed in experiments. Insisting on linear superposition and unitary time evolution leads to interpretations of quantum mechanics such as many worlds\cite{everett1957relative}, consistent histories\cite{griffiths1984consistent}, or the  relational interpretation \cite{rovelli1996relational}. These interpretations are however difficult to swallow for many people and are criticized in particular in connection with quantum measurement\cite{adler2003decoherence,schlosshauer2005decoherence,ellis2014evolving}. 

Another problem of the mentioned "derivations" of QSM from QM (and in general of decoherence theory) is that of justifying the assumptions required for obtaining a diagonal density matrix for the system. The calculations done to this purpose always include assumptions such as statistical independence and "typicalness" that are foreign to a deterministic theory\cite{kastner2014einselection,drossel2015relation}.

The third problem is that the amount of information required  in order to specify a many-particle wave function and to calculate its time evolution increases exponentially with the particle number, so that it will forever be impossible to test empirically the existence of such wave functions beyond  simple systems. Walter Kohn, who won the 1998 chemistry Nobel prize for developing density functional theory, warned us not to take the concept of a wave function too far \cite{lecture1999electronic}: ``In general the concept of a many-electron wave function $\Psi(r_1, \dots, r_N)$ for a system of $N$ electrons is not a legitimate scientific concept, when $N > N_0$, where $N_0 \simeq 10^3$. I will use two criteria for defining `legitimacy': a) That $\Psi$ can be calculated with sufficient accuracy and b) can be recorded with sufficient accuracy''. 

Due to these problems, the present paper explores a different avenue for connecting SM and QM. Instead of the reductionist approach that aims at  harmonizing SM and thermodynamics with a deterministic, unitary time evolution on the microscopic level \cite{callender2001taking}, we will argue that SM is not merely empirically adequate but that its stochasticity and irreversibility reflect true features of nature. If this is correct, SM can be used to identify the limits of unitary quantum mechanics. Thus, turning the reductionist agenda  and the title of Callender's paper "Taking thermodynamics too seriously" \cite{callender2001taking} around,  this paper is about "taking quantum mechanics too seriously".  

In the following section, we will consider the concepts, methods, and calculations used for thermalized many-particle systems. By studying ten different aspects of or approaches to thermalized systems, we will show that they require two important features: stochasticity and  a limited spatial extension of the wave function of a particle. If physics shall reflect at least approximately the features of reality, we have to conclude that in thermalized many-particle systems there are limits to the extension of a wave function, to its deterministic evolution, to linear superposition, and to entanglement.

Most of the following is formulated for a simple model system, namely a diluted gas of atoms with short-range interactions, which behaves in very good approximation like an ideal gas.


\section{Ten ways in which the physics of thermalized systems limits deterministic, unitary quantum mechanics}
\label{sec:tenreasons}

\subsection{Molecular dynamics simulations use localized atoms}

Molecular dynamics (MD) simulations are successfully used to characterize the structure and dynamics of systems that consist of many atoms at a finite temperature \cite{tuckerman2000understanding,marx2000ab}. Such simulations evaluate the motion of atoms and molecules based on the forces between them and describing the system at least in some respects classically. Temperature is taken into account by coupling the system to a ``thermostat'' that extracts and adds energy in such a way that a Maxwell-Boltzmann distribution of velocities is obtained. The majority of methods use Newton's equations of motion to calculate the motion of the nuclei. In purely classical simulations, molecules are represented as a collection of point masses and charges with restrictions on their relative positions, and forces are effective forces (e.g., van der Waals) obtained from comparison with empirical data or from quantum mechanical calculations. This type of simulations gives very good results for structural relaxation times and concentration profiles in liquids \cite{klameth2013structure}, for many biological processes such as transport through pores in biological membranes \cite{gumbart2005molecular}, and for protein folding \cite{karplus2005molecular}. When, however, the formation and breaking of bonds, the polarization of atoms or molecules, or excited states shall be taken into account, the quantum mechanical properties of the electrons must be considered, employing ab initio MD simulations. For given positions of the nuclei, the electronic structure of atoms and molecules is calculated using quantum mechanics. The motion of the nuclei is then calculated classically based on the force fields resulting from the electronic structure, and the electronic structure in turn is recalculated based on the changed positions of the nuclei. When quantum mechanical properties of nuclei become important, for instance with proton transfer processes that involve tunnelling,  the Feynman path integral formalism of statistical
mechanics is used to describe the nuclei. In this formalism, the partition function of one particle is written as
$Z(\beta)=\int dx \langle x | e^{-\beta H} | x \rangle $
with the quantum mechanical Hamiltonian $H=T+U$ (with $T$ being here the kinetic energy and not the temperature as in all other equations). This partition function can be rewritten such that it is identical to that of  a classical  harmonic chain that is closed to form a ring, in the external potential $U$. The typical size of this ring polymer is of the order of the thermal de Broglie wave length $\lambda_{th}=h/\sqrt{2\pi m k_B T}$. For $N$ particles, this partition function is extended such that the particles are treated like $N$ such ring polymers, with Boltzmann statistics. This latter step is an approximation that assumes that the $N$ particles are distinguishable and not entangled with each other. 

In all these different methods for performing MD simulations, the atoms are localized: they are points for classical MD simulations, they have the extension of the electronic shell for ab initio simulations, and the extension of the thermal wavelength for path integral ab initio simulations. In none of these approaches are the atoms or molecules entangled with each other, even though the time evolution of the system according to the full Schr\"odinger equation for all particles would yield such an entanglement. The time evolution of a $N$-particle system at finite temperature is thus described by the deterministic motion of objects that are localized within a small spatial region, with added stochastic terms due to the coupling to a heat bath. The microscopic picture of thermalized systems suggested by MD simulations involves only a limited range of quantum superposition and unitary time evolution. The idea that the atoms of a thermalized gas are best viewed as localized wave packets will be supported by the next subsections. 

\subsection{In thermal equilibrium the thermal wave length sets the length scale for quantum effects}

In statistical mechanics courses, sometimes the following quick derivation of the Bose-Einstein condensation temperature of an ideal Bose gas is given: Bose-Einstein condensation happens when the density of the gas becomes so large that the distance between atoms becomes of the order of the thermal wave length. This leads to 
$(V/N)^{1/3} = \lambda_{th} = h/\sqrt{2\pi m k_B T} $ and consequently $T = (N/V)^{2/3}h^2/(2\pi m k_B)$, which is apart from a numerical factor of the order 1 identical with the condensation temperature derived from the fully-fledged calculation. The thermal wave length emerges naturally in calculations of the canonical partition function of an ideal gas. Its order of magnitude can also be estimated without performing calculations from the equipartition theorem  $\frac 3 2 k_BT=E=h^2/(2m\lambda_{th}^2)$. The ratio $\lambda_{th} ^3/V$ is often called the \emph{quantum concentration}, see for instance the textbook \emph{Thermal Physics} by Kittel and Kroemer \cite{kittel1980thermal}.  This quick derivation of the Bose-Einstein condensation temperature is justified by arguing that the atoms can be represented as wave packets of the extension of the thermal wave length. When the density is so high that the wave packets overlap, atoms tend to go into the same quantum state. In the opposite case that the density is so small that the wave packets are not in contact for most of the time, the gas can be approximated as a classical ideal gas of well localized atoms. Similar arguments can be made for fermionic gases: When the density is so large that all the wave packets touch each other, it cannot be further increased due to the Pauli principle, leading to a Fermi temperature that is up to a constant factor identical with the Bose-Einstein condensation temperature. Again, in the limit of very low density, quantum effects become completely irrelevant, and the Fermi gas can be treated like a classical gas. For thermally equilibrated fermions as well as for bosons, the specific quantum mechanical effects become thus only important when the concentration is not small compared to the quantum concentration. 

This argument thus suggests that at sufficiently high temperatures atoms in a gas can be described as classical particles with a velocity distributed according to the Maxwell-Boltzmann distribution. This raises a puzzle: if all the wave packets that represent these atoms evolved according to the Schr\"odinger equation, scattering processes between them would cause them to become delocalized. Then the description as classical particles (or localized wave packets) would fail. The particles will remain localized only if there are limits of validity of the unitary time evolution of the Schr\"odinger equation. We will discuss this process of localization in more detail in the next subsection.

The localization of particles is necessary if there shall be a continuous connection between the description of a thermalized gas by quantum mechanics and its description by classical mechanics. While it is widely acknowledged that quantum mechanics shows the limits of classical mechanics by limiting the precision of points in phase space due to the uncertainty relation, the considerations of this and the previous subsection show that the opposite might also be true: if the classical description shall be approximately correct at densities far below the quantum concentration, the wave functions that describe the atoms of the gas must become localized within a small distance. Otherwise, there would be no justification for the considerations made in the foundations of classical statistical mechanics.  In fact, some authors point out that quantum mechanics always depends on the classical world \cite{weinstein2001absolute,drossel2015relation}.

\subsection{The concept of entropy implies random transitions between a finite number of different states}

Boltzmann's entropy formula is $S_B=k_B\ln \Omega$, where $\Omega$ is the number of different microstates that represent the same macrostate. While this definition requires some kind of coarse-graining and the choice of an appropriate measure of phase space in classical SM,  its evaluation in QSM requires simply the counting of quantum states (eigenfunctions) within a small energy interval. The same result is obtained by evaluating the Gibbs entropy $S_G=-k_B\sum p_k \ln p_k$, since in a macroscopic finite-temperature system almost all states lie within a small energy interval such that the probabilities $p_k$ can taken to be identical $p_k=1/\Omega$. 

A popular way of explaining entropy consists in saying that it is a measure of the information required to specify the microstate of a system if only the macrostate is known. If this information is given in bits, then one obtains the entropy by multiplying the number of bits with $k_B \ln 2$. 

While probably all researches would agree so far, there is considerable disagreement about how to interpret the probabilities occurring in Gibbs' expression \cite{frigg2011entropy}. A simple and straightforward interpretation consists in stating that these probabilities reflect a true stochasticity of the system, which means that the system moves in a not fully deterministic manner through its different microscopic states. The fundamental axiom of statistical mechanics that in a closed system in equilibrium all accessible microstates are equally probable, can naturally be grounded in such an idealization of the system. 
The second law of thermodynamics follows immediately from this idea. When a system is not closed but in thermal contact with another system, all statistical properties are obtained by assuming that there are stochastic transitions between different states, leading to a random distribution of the energy and the particles over the two systems. 

In contrast, to those who consider the system as following on a microscopic level a deterministic time evolution according to the Schr\"odinger equation,  the usage of probabilities is not due to an inherent stochasticity, but merely to our ignorance of the precise state of a system \cite{jaynes1957information}. On a microscopic level, a Laplacian demon could calculate the state of the system at later times based on an initial state, and the entropy would vanish. Such a subjective view of probabilities can only lead to agreement with empirical observation if the precise microscopic state is  irrelevant for the value of any observable quantity. This principle was framed by Jaynes in the form of the principle of maximum entropy: the probabilities take those values that maximize entropy given all macroscopic information about the system. The research agenda of eingestate thermalization \cite{deutsch1991quantum,srednicki1994chaos}
is intended to justify this principle. 

This widespread view has several problems: First, a deterministic theory cannot get rid of the randomness inherent in the probabilities of SM, but it simply moves it to the initial conditions, which are described as being "random" or "typical" or as having no hidden correlations that would lead at later times to a particular behavior. 
In order to argue that a system shows ergodicity or eigenstate thermalization, one must assume that the initial state of the system is not in one of the special states the time evolution of which deviates from the ``typical'' behavior. However, this is a strong assumption that is definitely not correct when applied backwards in time, since the present equilibrium states have come from very ``improbable'' initial states.  Furthermore, this strong assumption amounts to saying that a random state and a stochastic time evolution of the system are in all respects sufficient for describing  the system. Then, this should be taken as a valid description, without postulating that there is a (completely irrelevant) underlying deterministic dynamics. 

Second, in order to fully specify the state of a deterministic system, the number of bits required is infinite and not finite. In a quantum mechanical description, a thermalized system is not in an eigenstate of the Hamiltonian, but in a superposition. The expansion coefficients that specify this superposition are real numbers, which are given by infinitely many bits. The limited number of states occurring in the mathematical expressions for entropy suggests again that only a limited precision is relevant for capturing all those features of the system that are accessible to empirical observation and measurement. However, with a limited precision the information about the far past as well as about the far future is not contained in the present state. In this way, a limited precision is closely tied to stochastic dynamics. 

Therefore, let us take the stochasticity suggested by a nonzero entropy seriously. In fact, this stochasticity can be related to the suggestion of the first two subsections that a dilute gas consists of $N$ wave packets of a volume of the order $\lambda_{th}^3$. If these wave packets did evolve deterministically according to the Schr\"odinger equation, they would broaden with time due to dispersion while moving freely, and they would become entangled with other wave packets during scattering events. In particular during head-on collisions the wave packets would pick up a broad angular distribution.  Thus, in order for the wave packets to remain compact, they must become localized again after scattering, and this means that there occurs some kind of  ``collapse''.

\subsection{Thermalized systems are extensive/have statistically independent subparts}
Thermodynamics of homogeneous systems is extensive: if identical systems are combined to form a larger system, the extensive state variables $V$, $N$, $S, \dots$ and the thermodynamic potentials $E$, $F, \dots$  are the sum of those of the parts. The intensive variables $p$, $\mu$, $T, \dots$ do not change. This means that the state of each system is not changed in any relevant way when the systems are combined. Conversely, parts of a larger system do not depend in any relevant way on the neighboring parts. In statistical mechanics, the statistical independence of the parts of a system is an important precondition for deriving the probability distributions associated with the different statistical ensembles and for deriving statements about the size of fluctuations. This is stated most clearly on the first pages of the textbook on statistical mechanics by Landau and Lifschitz (in \S 2 of \cite{landau2013statistical}): The probability that a (sub)system in thermal equilibrium  assumes a certain configuration must be taken to be independent of the configurations of the neighboring (sub)systems. Then the extensive thermodynamic variables are simply sum variables over all subsystems, and their variance is the sum of the variances within the subsystems. In fact, as stated by Landau and Lifschitz, this gives an interpretation of the ``ensembles'' introduced by Gibbs as an appropriate description of a single large system by interpreting the subsystems of the large system as constituting the ensemble. Such a view of the Gibbs approach solves the interpretational problem of reconciling the ensemble view with applying SM to single systems mentioned by Frigg\cite{frigg2010field}. 

Now, in order to justify statistical independence of subsystems, one must argue that correlations between subsystems decay fast and are not observable in thermal equilibrium. The arguments of the first three subsections present a plausible justification for statistical independence, as they limit the range of quantum superpositions and of deterministic time evolution, leading to a limit memory of the past and therefore a limit survival time of correlations.

\subsection{Statistical mechanics describes the properties of single systems}

The view suggested in the previous subsection that the Gibbs ensemble should be interpreted as the subsystems of one large system is helpful when trying to understand why statistical mechanics captures correctly the properties of single macroscopic systems. The quantum statistical description of such macroscopic systems  uses the concept of a mixed state, characterized by a density matrix that is diagonal in the basis of the energy eigenstates. For a microcanonical ensemble, the density matrix is proportional to the unit matrix and is thus independent of the chosen quantum mechanical basis. In each basis, all states occur with equal probability. In particular, a wavelet basis that represents atoms as localized wave packets, is equally suitable as the usually chosen basis of energy eigenstates. The standard interpretation of the density matrix is that of an ensemble of systems, with the probabilities giving the proportion of systems in the different quantum mechanical states. In this interpretation, the description of the system as in a mixed state is again due to our ignorance of the full microscopic details. The probabilities are again subjective probabilities. This is in striking contrast to the fact that the mixed density matrix is so successful at describing the properties of single macroscopic systems, for instance when correlation functions or thermodynamic variables are calculated. A mixed state should therefore be taken as capturing something correct about an individual system, namely that a macroscopic system is not in a pure quantum state and cannot be described by a $N$-particle wave function. Let us therefore discuss which properties a system should  have if it shall be described correctly by a the density matrix of a mixed state. First, let us take up the suggestion of Landau and Lifschitz to imagine the system as composed of statistically independent parts. For the sake of simplicity, let all parts be of identical size. If the density matrix is taken to be that of a part, all the parts together form an ensemble. Then, the probabilities occuring in the density matrix can be interpreted as probabilities for the different states of the different parts. Now add to this the idea that there are random transitions between the states of each part. Then the assumptions that the parts are statistically independent from each other and that the entries of the density matrix represent objective probabilities become justified, and it follows that a mixed density matrix is a good description of the system. 

As an aside, let us mention that this interpretation of the mixed state avoids the puzzle of ergodicity in classical statistical mechanics: the time it takes the system to visit each cell in phase space with a deterministic time evolution is incredibly much longer than the life time of the universe, which means that ergodicity (or quasi-ergodicity) cannot explain the rapid approach to equilibrium. In contrast, with a stochastic evolution each state of the system can be reached within a short time: Let us take $10^{23}$ atoms of a gas at ambient temperature and pressure. The particle density is of the order $25 \cdot 10^{24}/m^3$, which means that the mean distance is of the order 30 \AA. If an atom has the size of 1\AA, it will move about 1$\mu m$ between collisions. With a speed of the order of $10^3 m/s$, there will be of the order of one collision per $ns$ per atom. Assume that the number of possible states to which a wave packet can become localized after a collision is of the order of 10, then there are of the order of $10^{10^{23}}$ states that the system can reach within $1ns$. The total number of states available to the system is of the order of $(V/N\lambda_{th}^3)^N$, which is of the order of $10^{10^{24}}$. Such a large number of states can thus be reached within 10 collision events per atom, i.e., within 10$ns$.

\subsection{A true equilibrium has forgotten the past}

A thermalized system is in equilibrium. A thermal equilibrium state contains no information about the initial conditions, and it satisfies detailed balance. This means that its time evolution can in no way be distinguished from the time-reversed situation. The approach to equilibrium is an irreversible process, during which entropy increases until it is maximum in equilibrium. 

This is the general understanding of equilibrium. A unitary, deterministic time evolution is fundamentally different, since it contains the full information about the initial state, and since the "equilibrium state" is therefore not invariant under time reversal. In order to reconcile it with a behavior that resembles an approach to equilibrium, a coarse-grained view on the system based on relevant observables, combined with "typical" initial states is required \cite{gogolin2016equilibration}. 
In contrast, the usual way to model the approach to equilibrium involves equations that are not invariant under time reversal, such as diffusion equations or equations with friction terms. A beautiful simple theory that leads to an equilibrium state for a gas is the Boltzmann equation, which can be found in many textbooks on statistical mechanics. In the absence of external forces, it has the form
\begin{eqnarray} &&\frac{\partial f(\vec p,\vec q,t)}{\partial t} +
\dot {\vec q}\cdot \frac{\partial f(\vec p,\vec q,t)}{\partial \vec q}
= \int d^3 p_2 \int d^3 p_3 \int d^3 p_4 \nonumber \\
&& W(\vec p, \vec p_2; \vec p_3,
\vec p_4) \left[f(\vec p_3,\vec q,t)f(\vec p_4,\vec q,t) -f(\vec
  p,\vec q,t)f(\vec p_2,\vec q,t)\right]\, , \nonumber
\label{boltzmann}
\end{eqnarray}
where $f(\vec p,\vec q,t)$ is the particle density in the 6-dimensional phase space spanned by the momentum and position coordinates of a particle. The right-hand side describes the momentum changes due to collisions, with the function $W$ depending on the scattering cross section. The collision term neglects correlations between the momenta of different particles, and this leads to an irreversible behavior of the system, with the function 
\begin{equation}
 H(t)=\int d^3 p \int d^3 q f(\vec p, \vec q, t) \ln f(\vec p, \vec q,t)
\end{equation}
decreasing in time until the equilibrium state is reached, which satisfies the detailed balance condition
$$f(\vec p,\vec q,t)f(\vec p_2,\vec q,t) = f(\vec p_3,\vec q,t)f(\vec p_4,\vec q,t)\, .
$$

The neglection of correlations between the momenta of different particles is another form of the assumption of statistical independence mentioned earlier. In fact, statistical independence means that the past history is not important for the present behavior, and that there are no subtle interdependencies that arise through the dynamical history. 

A similarly simple quantum mechanical analogon of the Boltzmann equation does not yet exist. The quantum theory that comes closest to Boltzmann's theory of relaxation to equilibrium is probably that of quantum diffusion \cite{caldeira1983path}. Here, the quantum description of the motion of a particle that is coupled to many external degrees of freedom is given. Not surprisingly, the calculations that eventually give the Langevin equation of Brownian motion include assumptions about a fast relaxation of the correlations of the external degrees of freedom which are therefore taken to be statistically independent on the considered time scales. This is the equivalent of the neglection of correlations in the Boltzmann equation. Subsection \ref{openquantumsystems} on open quantum systems discusses in more depth the approximations that are made when performing quantum calculations for systems that are in contact with a heat bath.

To conclude, the usual way of modeling the transition to equilibrium involves irreversible equations. The underlying microscopic picture from which such irreversible equations can be obtained is stochastic with a limited memory of the past. This fits nicely together with the previous subsections. Accepting that there are limits to a deterministic, unitary time evolution thus appears unavoidable when QM shall be reconciled with SM. 

\subsection{Spontaneous symmetry breaking in a phase transition is a stochastic event}

When a system undergoes a (symmetry-breaking) phase transition, it ``chooses'' spontaneously and stochastically the new, symmetry-broken state into which it goes. Such a transition that is accompanied by the spontaneous breaking of a symmetry is incompatible with a unitary time evolution. A unitary time evolution that starts from a state that obeyes the symmetry under consideration and that evolves according to a Hamiltonian that also displays this symmetry must go to a final state that also has this symmetry. The final state therefore must contain all possible broken-symmetry outcomes with equal weights, and not just one of them. By taking the trace over environmental variables, such a final state would be represented by a mixed density matrix. 

Let us focus in the following on a discrete symmetry, such as the up-down symmetry in the Ising model.
Even if the Hamiltonian does not strictly preserve the symmetry that is relevant for the phase transition, a unitary time evolution of a quantum system is incompatible with the spontaneous symmetry breaking occurring in a phase transition.
In order to make this concrete, let us perform a Gedankenexperiment and construct a toy system, consisting of an Ising model (with the $z$ component of neighboring spings being coupled) in contact with a heat bath, which consists in a ``real'' system of the phonons of the magnetic material, but for the sake of the argument, we can also imagine it as the gas of the previous sections. Let the total energy of the system be such that the maximum entropy state is one with a broken symmetry in the Ising model, with the magnetization being $+M$ or $-M$ (with respect to the $z$ direction). Such a state of broken symmetry has a lower internal energy than a disordered spin system, and the energy freed by the ordering can go into the heat bath, where the entropy increases. (This is equivalent to saying that the spin configuration minimzes the free energy of the spin system.) Let us choose an initial state where the spins are oriented at random, and where the state of the spin system can be written as a product of the states of all spins. In the following, I will show that the assumption that the magnetic system orders to the $+M$ or $-M$ state while the time evolution of the total system is unitary leads to a contradiction. Let us start with a randomly chosen state of the spin system, and a random state of the environment. Assume that under unitary time evolution the spin system evolves to the $+M$ state. Now prepare again the same initial state of the total system, but with one of the spins reversed. If the unitary time evolution goes again to the $+M$ state, then take again the same initial state, but with an additional spin reversed. Eventually, an initial spin state will be generated that leads to the $-M$ state. Then we take a linear superposition (with equal weight) of the last spin state and the one before as initial state of the spin system, again combined with the same environmental initial state. The state of the spin system is again a product state, but with the spin that was reversed last now pointing in a direction perpendicular to the ones before. Due to the linearity of the Schr\"odinger equation, this state must now evolve to a zero magnetization, in contradiction with our assumption that every initial state that is a product state of spins ends up with a magnetization $\pm M$. 

A stochastic time evolution does not face this problem of a unitary time evolution. In fact, computer models based on transition probabilities between states are very successful at modelling phase transitions. Such Monte-Carlo approaches with Glauber or Metropolis dynamics capture correctly the equilibrium properties for instance of an Ising system. 
This suggests that the idea that a thermalized system undergoes truly stochastic transitions between its different states captures something correct about reality. 

As a side note, the measurement process faces exactly the same problem as a thermodynamic phase transition: linear superposition is in contradiction with the observation that a measurement always gives a specific outcome. The spontaneous selection of one measurement outcome is analogous to the spontaneous symmetry breaking in a phase transition. In fact, the system of the previous section can be used as a measurement apparatus: We only need to include an additional coupling of  the Ising system to an external spin 1/2 that is sent into the system and the $z$ component of which shall be measured: if the coupling is in a suitable range of values, the Ising system will go to the $+M$ configuration if the spin is prepared in the $+$ eigenstate of $\sigma_z$, and to the $-M$ configuration if the spin is prepared in the $-$ eigenstate (for a discussion of an explicit model of this type, see \cite{allahverdyan2005phase}). The final magnetization of the Ising system can be used as a pointer from which the measurement result is read off. When the spin is prepared with an orientation in the $+x$ direction, time evolution according to the Schr\"odinger equation leads to a linear superposition of the two previous outcomes and not to a unequivocal $+M$ or $-M$ result, each with probability 1/2.

\subsection{The relation between fluctuation and dissipation in thermalized systems  presupposes equilibrium and forgetting of the past}

The response of a thermalized macroscopic system to an external influence, for instance a field, is very reproducible and is closely related to the fluctuations of the inner degrees of freedom of the system. The mathematical expressions for these relations are called  fluctuation-dissipation (or fluctuation-response) relations. Such relations hold as long as the external influence is not too strong so that thermal equilibrium is not destroyed. In this case linear response theory is sufficient to describe the system.
Examples are the relation between the damping constant and the diffusion coefficient of a Brownian particle, the relation between compressibility and density fluctuations in a fluid,  the relation between specific heat and entropy fluctuations, or the relation between magnetic susceptibility and the fluctuations of the magnetization. The derivation of the fluctuation-dissipation theorem (see for instance chapter 4 in \cite{schwabl2005advanced}) is never done starting from deterministic, time-reversible microscopic equations. Instead, the expression for the fluctuations is evaluated using the canonical or grand canonical density operator, and the response or dissipation is calculated using "causality", which means that the system responds to the external influence after the influence occurs and not before. This leads for instance to the Kramers-Kronig relation that establishes a close connection between the real and the imaginary part of the susceptibility (or response function).  

The physical, intuitive explanation for fluctuation-dissipation relations is that the energy that is put into the system by the external influence in a non-random way, becomes randomly distributed in the system. The underlying picture is that the inner degrees of freedom perform random fluctuations by which energy is randomly exchanged between them. When an external field or another external influence is applied,  energy is put into the system via those degrees of freedom that couple to this influence, and is then redistributed between all degrees of freedom that participate in the thermal equilibrium in order to generate the new equilibrium in the presence of this influence. It is essential that time reversibility is broken: "causality" means that there is "first the cause and then the effect". As mentioned above, this fact is explicity implemented in the derivations of fluctuation dissipation relations. However, such a concept of causality is totally foreign to a time-reversal invariant microscopic theory where the present state contains implicitly all future states as well as all past states. In contrast, a system that responds in a reproducible way to external influences must be in a state of indifference that has no memory of the past (or at least none that is relevant for the response) and no intrinsic goals for its time evolution. Furthermore, it must "forget" the influence soon after it subsides. 
A random exchange of energy between the inner degrees of freedom that leads to equilibration achieves all this. 

Just as in the previous subsections, a straightforward interpretation of the physical calculations done for thermalized systems suggests that these systems have a truly stochastic dynamics and have a limited memory of their past. 

\subsection{The photons of black-body radiation follow a continuous, universal spectrum and are emitted locally}

Every macroscopic finite-temperature system emits thermal radiation from its surface, which obeys the Planck spectrum of black-body radiation.  This radiation has universal properties that do not depend on the particularities of the system. The spectrum depends only on temperature, and the radiated power is proportional to the surface area. 

Now, the emission of a photon requires a quantum transition between different states in the system. Since the Planck spectrum is continuous, these transitions cannot be electronic transitions within atoms or molecules. Rather, in a gas, the thermal photon emission occurs via bremsstrahlung, and in a crystal via phonon interactions. In all cases, the electromagnetic interaction between the atoms  of the system is relevant. In order to generate a spectrum that is continuous and universal, the quantum states between which the transitions occur must be sufficiently "random" or "generic". Furthermore, in order to generate and an intensity that is proportional to the surface area, these transitions must occur independently within regions of a sufficienty small size. If the system was described by one comprehensive wave function of all particles, one would again need to argue that with respect to the transitions relevant for photon production this wave function can be considered as sufficiently random and the transitions as occurring within localized regions. This, however, has never been done. Instead, the explicit modeling of the interaction between phonons and photons is done by using quantum field theory. In these calculations, phonons are modeled as pretty well localized (quasi-) particles that interact locally. This can be justified only if there is a limited range of quantum superpositions and a limited memory of past interactions. Thus  the properties of black-body radiation and the theoretical treatment of the interaction between phonons and photons suggest again that the time evolution of wave functions according to the Schr\"odinger equation has a limited scope.  

\subsection{The theory of thermalized open quantum systems uses the Markov assumption and a product ansatz for the combined state of system and environment}\label{openquantumsystems}

Thermalization of a system means that it is in equilibrium with an external macroscopic environment that imposes a temperature on the system. 
Since no system can be completely isolated from the rest of the world, the theory of open quantum systems has been developed to take into account the influence of the environment when describing a system (see for instance \cite{breuer2002theory} for a good textbook). Tracing over the degrees of freedom of the environment leads to a quantum theory for the system alone that is no longer deterministic nor time reversal invariant. 

Now, these "derivations" are usually understood as demonstrating that the irreversible, stochastic features of SM can be derived from QM. However, a close look at these "derivations" reveals that all of them must presuppose in one way or another what they conclude in the end. On the one hand, all of them use product states as initial states. However, based on a deterministic, unitary time evolution there is no way to justify such a product state where the system is initially not entangled with anything else. Furthermore, all "derivations" make repeated assumptions of uncorrelatedness of the degrees of freedom of the environment. This is for instance done in form of the Markov assumption when deriving Lindblad equations, or in form of zero averages when arguing in decoherence theory that the non-diagonal elements of the reduced density matrix vanish.

 Let us take as an example the simple model for decoherence given in \cite{cucchietti2005decoherence}, where a thermal environment is coupled to a spin-1/2. The environment is modeled as consisting of 2-state systems, which means that each ``atom'' of the environment is formally equivalent to a spin-1/2. Now, in order to obtain a diagonal reduced density matrix for the spin, the coupling constant of the environmental ``atoms'' to the spin are assumed to be random, and similarly the phases of the ``atoms'' that arise during time evolution are assumed to be uncorrelated. This is in fact an assumption of statistical independence, which means  that the argument for decoherence is circular: it presumes statistical independence in order to ``derive'' it subsequently \cite{kastner2014einselection}.  

Moreover, as mentioned in the introduction, 
decoherence theory has an additional problem as it still gives a - incoherent - superposition of all the possible time evolutions of the system and not a single realization as in classical physics \cite{adler2003decoherence}.

\section{Discussion}

The arguments presented in the previous section suggest a microscopic view of thermalized systems that is based on how physical calculations for thermalized systems are actually done. This view includes two general features that are in contradiction with a deterministic, unitary time evolution and linear superposition, which are features of nonrelativistic quantum mechanics. 

First, there is a limited range of linear quantum superposition. This feature took different shape in different subsections.  In the first two subsections on MD simulations and the length scale for quantum effects, it took the form that the atoms of the gas are described between scattering events as localized wave packets of an extension of the order of the thermal de Broglie wavelength.  In the 4th subsection, it occurred implicitly in form of the statistical independence of subparts, and in the 5th subsection it allowed the interpretation of Gibbs' ensembles as subsystems of one large system. Statistical independence was also invoked in the subsections on black-body radiation and open quantum systems, where it took the form of a product ansatz.  

Second, the system has a stochastic contribution to its dynamics which is responsible for forgetting the past and approaching equilibrium. This features was essential in the subsections on entropy, equilibrium, phase transitions, and fluctuation-dissipation relations. 

It appears impossible that all the features of thermalized systems mentioned in the previous ten subsections can also be derived from a deterministic, unitary time evolution of a many-particle wave function according to the Schr\"odinger equation. As already mentioned, no calculation that starts with a deterministic theory can proceed without introducing stochasticity in the form of "typical" or "random" initial states and a limited range of quantum superposition by postulating initial product states or assuming a fast decay of correlations. The problem that the foundations of SM require stochasticity is already known from classical SM. Both types of approaches try to eliminate stochasticity by moving it back to the initial conditions when they postulate "random" or "typical" initial states. As mentioned by Gisin\cite{gisin2016}, there is no way to empirically distinguish stochasticity in the initial state from a a time evolution that is based on states that possess limited precision and has thus a stochastic component. This means that for all practical purposes stochasticity is a relevant feature of the microscopic time evolution that underlies statistical mechanics. In several respects, the problems of foundations of statistical mechanics are much worse in QSM than in classical statistical mechanics. While in the foundations of classical statistical mechanics the microscopic description of the system is given by the positions and momenta of the order of $10^{23}$ particles, the specification of the microscopic state in quantum mechanics requires of the order of $\exp(10^{23})$ complex numbers, which are the coefficients that specify the many-particle wave function in a chosen orthonormal basis, for instance a product basis of one-particle states. As mentioned in the Introduction, there is no hope to measure such a wave function even with a modest precision when dealing with many particles. For systems that are far away from the ground state or any other eigenstate, such as finite-temperature systems, it is furthermore impossible to generate exact copies of the same system. But in order to measure a wave function, it is necessary to use many copies of it. All this means that a many-particle wave function in a thermalized system is not an empirically testable quantity. Even worse, it leads to straightforward contradictions: The time evolution of a thermalized system embedded in an environment according to the Schr\"odinger equation leads to entangled states that consist of the linear superposition of an exponentially increasing number of products of system states with environmental states. However, in situations where the superimposed states differ in some macroscopic observable, as for instance due to symmetry breaking at a phase transition, only one of these superimposed states is found. This is of course the fundamental interpretational problem of quantum mechanics, which is more familiar from the measurement process, but it is equally viral in QSM. There is also a close connection between the measurement process and the arguments presented in the previous section about the thermalized gas consisting of localized wave packets: localization of the measured particle is an essential part of a measurement process. Now imagine an additional atom entering the thermalized gas, and assume that this atom was prepared with a sharp momentum, i.e. as a plane wave. When this atom becomes part of the gas and is eventually in equilibrium with the rest of the gas, it must also become a localized wave packet, just as the other atoms of the gas. The ``problem'' of the measurement process is thus connected in two ways to the ``problem'' of quantum foundations of statistical physics: First, both involve a macroscopic number of degrees of freedom at finite temperature (in the measurement process these are the internal and/or environmental degrees of freedom of the measurement device), second, both show a spontaneous symmetry breaking where one of the possible outcomes is chosen, while unitary time evolution leads to a superposition of the possible outcomes (in statistical physics this occurs for instance at a phase transition). 

Apart from all the problems when dealing with QSM, the belief in a deterministic unitary time evolution of a wave function is questionable also on other grounds: The Schr\"odinger equation is not exact, and we have no reason to assume that any of our physical theories is exact. The history of classical mechanics has taught us an important lesson in this respect: for more than 200 years, it was generally believed that Newtons's laws provide an exact description of nature and are valid even under circumstances where they had not been tested: on the atomic scale, at high velocities close to the velocity of light, at cosmic distances. The advent of the theory of relativity and quantum mechanics has revealed that Newton's laws are only an approximation to reality, with a limited range of applicability. Nevertheless, these laws are still extremely good and useful for many purposes. This experience should make us open to the idea that probably none of our theories is exact. As far as the Schr\"odinger equation is concerned, we know its limits of applicability: It ignores relativistic effects and in particular the creation and annihilation of particles. Furthermore, a description of a thermalized gas by a many-particle Schr\"odinger equation neglects the presence of photons, which are emitted and absorbed in the system and which obey a Planck spectrum. The electromagnetic interaction is treated classically when a system is described by a Schr\"odinger equation. While the Schr\"odinger equation is an extremely good description for many purposes, we have no reason to expect that it applies also to a system with $10^{23}$ particles at finite temperature. On the contrary, the points emphasized in this paper strongly suggest the Schr\"odinger equation (or any modification that preserves linear superposition and determinism) has limited validity in such systems.

\section{Concluding remarks}

This article criticized the reductionist view of thermalized systems that claims that the underlying microscopic reality is  a highly entangled (not just between the particles of the system, but also with the rest of the world!) many-particle wave function which evolves deterministically and still contains all information about the initial state. Apart from causing severe interpretational problems and being not empirically testable, this explanation requires ad-hoc assumptions about statistical independence that cannot be justified within a deterministic theory. 
In contrast,  this article recommends to take the area of physics that is best suited to describe a thermalized system seriously, namely statistical mechanics and thermodynamics. There is no reason apart from prejudice why we should not take these theories to be as close to reality as the many-particle Schr\"odinger equation, with each of these descriptions having their own limited range of applicability.

This, of course, raises many follow-up issues, some of which shall be mentioned in the following.

\subsubsection*{Relativistic quantum theories}

While everybody acknowledges that the Schr\"odinger equation is only approximately valid, it is more widely believed that a suitable relativistic quantum theory is an exact description of nature. Often, this is combined with the belief that relativistic time evolution is also deterministic and unitary and can be linearly superimposed. The discussion about the possible information loss in black holes\cite{blackhhole} is exactly based on such a belief. Based on the arguments of the present paper, information loss in black holes presents no problem but is to be expected due to the stochasticity of finite temperature systems. In fact, relativistic quantum mechanics cannot do without stochasticity and assumptions about uncorrelatedness.  The formalism of quantum field theory does not only operate with unitary time evolution, but it requires a preparation of the initial state (which is not entangled with the rest of the world) and the projection on a final state. These two steps introduce again a stochastic component (since preparation and projection involve measurement, which gives stochastically one of the possible results) and they destroy correlations because the initial state is taken to be a product state of the considered particle or system and of the system with which it will interact subsequently.  This means that quantum field theory cannot do without the two features that we have emphasized in the discussion of thermalized systems. 

Very general arguments why a linear, unitary time evolution must have limits of validity when considering compex systems are made by G.~Ellis \cite{ellis2012limits}.

\subsubsection*{Statistical interpretation}
Proponents of the statistical interpretation (also called ensemble interpretation) claim that this interpretation solves many of the problems mentioned above \cite{Ballentine}. This interpretation holds that the wave function does not describe an individual system but an ensemble of systems. It traces back to Max Born who suggested that the wave function gives the probability amplitude for measuring the particle at a given position. 
According to the statistical interpretation, quantum mechanics of pure states is a special case of a more comprehensive theory, which is quantum statistical mechanics. This interpretation does not encounter some of the problems mentioned in this paper, but it has other problems: With this interpretation, quantum mechanics is incomplete and cannot describe single systems. In addition, it has an internal inconsistency: The Schr\"odinger equation describes the time evolution of a ``probability distribution'', but there is no criterion to decide when the event described by this probability distribution happens. This means that there is no distinction between an ensemble of systems in which each system evolves forever according to the Schr\"odinger equation, and an ensemble where in each system a ``collapse'' to one measurement result happens. 
Furthermore, there is empirical evidence that a pure quantum state does in fact represent single systems \cite{pusey2012reality}.

\subsubsection*{The meaning of temperature}
Usually, temperature is understood to be the mean kinetic energy of the particles of a many-particle system at equilibrium. However, if arguments of this paper are correct, temperature plays an important role at determining the range of quantum interference. In this case, it must be more than just a measure of the kinetic energy of particles. In fact, there are various examples in the literature that hint at the role of temperature at determining length or time scales related to the quantum description of many-particle quantum systems. In the non-relativistic quantum theory of open systems, the so-called "thermal time" $\hbar/k_BT$ sets the time scale beyond which the dynamics of the system shows the Markov property \cite{weiss1999dissipative}. This time is of the order of the time that a wave packet with a kinetic energy of the order of $k_BT$ needs to cover the distance of the thermal wavelength. Based on the intuition developed in section \ref{sec:tenreasons}, this sets the time scale beyond which unitary time evolution according to the Schr\"odinger equation breaks down and wave packets become localized. Concordantly, theories that describe a spontaneous collapse of wave packets also include constants that correspond to a "temperature" \cite{bassi2013} and that limit the time scale over which unitary time evolution occurs. However, the proponents of such theories usually do not consider the physical temperature of many-body systems as the relevant temperature for the collapse. In quantum gravity, the concept of a "thermal time" also plays an important role \cite{rovelli1993statistical}. Since the theory of general relativity shows interesting parallels to thermodynamics \cite{jacobson1995thermodynamics}, and since gravitational fields are believed by some scientists to cause a collapse of the wave function \cite{penrose1996gravity}, the relation between thermodynamics and the limits of unitary time evolution discussed in this paper might have a far wider scope. 

\subsubsection*{Quantum mechanical chance is a collective effect}
The discussions done in this paper suggest that stochasticity arises in finite-temperature many-particle systems. However, it is often claimed that true chance enters the world on the microscale, through quantum events, for instance the radioactive decay of a nucleus or the spontaneous emission of a photon from an excited atom. In these situations, quantum chance appears to be related to one or a few particles. However, when considered more closely, this stochasticity arises in fact due to the interaction of an atom with the rest of the world (for instance the other atoms of a gas, or a measurement device) and is thus a collective effect at the interface to statistical physics. Chance is thus a context-dependent phenomenon. This view is in contrast to stochastic collapse theories \cite{ghirardi1986unified,ghirardi1990markov}, which include stochasticity already in the time evolution of an isolated particle.
 
 To conclude, the considerations in this paper touch upon some of the most important open questions of contemporary theoretical physics. In order to resolve the puzzles and inconsistencies in present theories at the interface between quantum physics and statistical physics, it is necessary to confront the limits of a deterministic, unitary time evolution.  

\acknowledgements{I thank Marina Cortez, Matteo Smerlak, Lee Smolin, and Steve Weinstein for helpful discussions about the relation between statistical physics, quantum physics, and cosmology during my sabbatical. This work was supported in part by Perimeter Institute of Theoretical Physics. Perimeter Institute is supported by the Government of Canada  through the Department of Innovation, Science and Economic Development and by the Province of Ontario through the Ministry of Research, Innovation and Science.}

\end{document}